\documentclass[pdflatex,sn-mathphys-num]{sn-jnl}

\usepackage{graphicx}%
\usepackage{multirow}%
\usepackage{amsmath,amssymb,amsfonts}%
\usepackage{amsthm}%
\usepackage{mathrsfs}%
\usepackage[title]{appendix}%
\usepackage{xcolor}%
\usepackage{textcomp}%
\usepackage{manyfoot}%
\usepackage{booktabs}%
\usepackage{algorithm}%
\usepackage{algorithmicx}%
\usepackage{algpseudocode}%
\usepackage{listings}%

\theoremstyle{thmstyleone}%
\theoremstyle{thmstyletwo}%

\theoremstyle{thmstylethree}%

\begin{document}

\title[Article Title]{Thin-film lithium tantalate for ultraviolet integrated electro-optic modulator

}


\author*[1]{\fnm{Chupao} \sur{Lin}}\email{Chupao.Lin@UGent.be}

\author[1]{\fnm{Patrick} \sur{Nenezic}}

\author[2]{\fnm{Arno} \sur{Moerman}}

\author[1]{\fnm{Konstantinos} \sur{Akritidis}}

\author[1]{\fnm{Tom} \sur{Vanackere}}  

\author[1]{\fnm{Simone} \sur{Atzeni}}

\author[1]{\fnm{Margot} \sur{Niels}}
\author[1]{\fnm{He} \sur{Li}}
\author[1]{\fnm{Valeria} \sur{Bonito Oliva}}

\author[1]{\fnm{Maximilien} \sur{Billet}}

\author[1]{\fnm{Bart} \sur{Kuyken}}

\affil[1]{Photonics Research Group, Department of Information Technology (INTEC),
Ghent University - imec, Ghent, Belgium}

\affil[2]{Department of Information Technology (INTEC), IDLab, Ghent University-imec, Ghent, Belgium}


\abstract{The realization of integrated, high-speed ultraviolet (UV) modulation is pivotal for the advancement of quantum information processing, portable atomic clocks, and secure solar-blind communications. While mature photonic platforms have facilitated sophisticated system-level integration across visible and infrared spectra, high-speed active modulation in UV remains with traditional bulk crystals. Consequently, a scalable integrated solution that simultaneously combines low insertion loss and extreme compactness with high modulation efficiency has remained challenging. Here, we report the first integrated UV electro-optic modulator on a thin-film lithium tantalate (TFLT) platform. By employing a compact lumped-electrode design, we achieve a record-low V$_\pi$L of 85 ~mV$\cdot$cm at 375~nm, providing an up to four orders of magnitude improvement in terms of bandwidth/V$_\pi$L over bulk technologies. The device demonstrates a robust extinction ratio of 22.7~dB, a low insertion loss of 1.3~dB, and a V\textsubscript{$\pi$} of 4.2V. Although the measured 3-dB bandwidth of 922 MHz is currently limited by photodetector performance, the small device footprint of 1.16 mm and electrode design of 200~$\mu$m indicate intrinsic potential for high-speed operation beyond 67 GHz which is confirmed by the electrical-to-electrical response. This work establishes TFLT as a disruptive platform for wafer-scale compatible active UV photonics, enabling the next generation of scalable quantum and communication systems.}


\keywords{Ultraviolet, Lithium Tantalate, Electro optic modulator, Pockels effect}



\maketitle

\section{Introduction}\label{sec1}

Precise manipulation of ultraviolet (UV) and visible light is fundamental to a diverse range of emerging technologies, as illustrated in Fig. 1. These include trapped-ion quantum computing and optical clocking \cite{ Corsetti2026,Mehta2020,Kwon2024}, high-precision UV spectroscopy \cite{Xu2024}, high throughput microscopy \cite{Lin2022,Lin2023}, solar blind communications \cite{Mo2025} and underwater optical communication \cite{Sun2018}. To date, UV optical architectures have relied predominantly on discrete bulk components, such as mirrors, lenses, and macroscopic modulators. However, this bulk-optical approach is inherently limited by poor scalability and high sensitivity to mechanical vibrations. Shifting toward an integrated photonic platform not only enables high-fidelity control over the amplitude, frequency, and phase of light but also fundamentally suppresses ambient control noise through a more robust and compact footprint \cite{Brown2021}. Driven by the need for telecommunications, platforms like silicon (Si) and silicon nitride (SiN\textsubscript{x}) have facilitated complex system-level integration at infrared wavelengths. Yet, this success has not translated to the UV regime, as the limited bandgaps of these materials fundamentally lead to prohibitive absorption in UV. The emergence of the photonic platform based on thin-film alumina (AlO\textsubscript{x}) provides a scalable architecture for large-scale passive photonic integration in UV \cite{Neutens2025,HENDRIKS2024}.  However, this platform lacks intrinsic gain media and a second-order ($\chi^2$) nonlinearity. Consequently, active functionality has been primarily restricted to thermo-optic (TO) phase shifters. While TO modulation is a well-established mechanism for tuning phase and intensity, its operation is fundamentally constrained by slow response times (on the order of 10s of kHz \cite{Wang2026}) and significant thermal crosstalk, both of which severely limit the attainable modulation bandwidth and circuit integration density. In the telecommunication bands, silicon photonics has demonstrated high-speed modulation up to the tens of GHz regime. In contrast, UV photonic platforms have not yet reached such maturity. Establishing a robust, high-speed active component library is essential to bridge this gap and unlock the full potential of UV photonic platform.
Realizing high-speed UV modulation remains a challenge, primarily due to the lack of low-loss thin films with significant electro-optic (EO) coefficients. While the platform of aluminum nitride (AlN) on sapphire offers a wide bandgap and a modest EO coefficient ($\sim$1~pm/V), its potential for UV photonic integrated circuits (PICs) is severely hampered by the lattice mismatch at the AlN-sapphire interface. This structural disparity induces high material absorption and scattering losses \cite{Liu2018}, fundamentally obstructing its integration into high-performance UV circuits. Alternatively, by leveraging the strong piezoelectric coefficient of aluminum nitride, Castillo et al. \cite{Castillo2026} recently integrated AlN piezo-optomechanical actuators with $\text{AlO}_x$ resonators. This architecture achieves intensity modulation of UV light at frequencies reaching the tens of MHz. However, the modulation bandwidth is inherently governed by the mechanical resonance frequencies of the actuator. Achieving gigahertz-rate integrated UV modulation is a critical milestone that will enable a suite of applications currently restricted to laboratory-scale, bulk-optical setups. For instance, a phase modulation frequency of 12.6 GHz is fundamental for addressing the hyperfine transitions of $^{171}\text{Yb}^+$ ions at 369.5~nm. Currently, high-speed UV phase modulation relies on bulk electro-optic crystals such as $\beta$-barium borate (BBO) and deuterated potassium dihydrogen phosphate (DKDP) \cite{qubig_am8_uv,qubig_pc_uv,qubig_pm11_uv}. However, their macroscopic geometry and wide electrode gaps necessitate half-wave voltages ($V_\pi$) on the order of kilovolts. This requirement imposes a severe bottleneck: the high-power RF drive needed to reach GHz frequencies induces substantial resistive heating, effectively restricting the sustainable intensity modulation rate to the hundreds of MHz regime. Despite the maturity of high-speed modulation in the near-infrared, a high-speed UV counterpart has remained challenging due to a main conflict: the difficulty of realizing a thin-film platform that simultaneously provides a high electro-optic coefficient, low propagation loss, and a wafer-scale compatible fabrication process.

\begin{figure}[!t]
\centering
\includegraphics[width=1\linewidth]{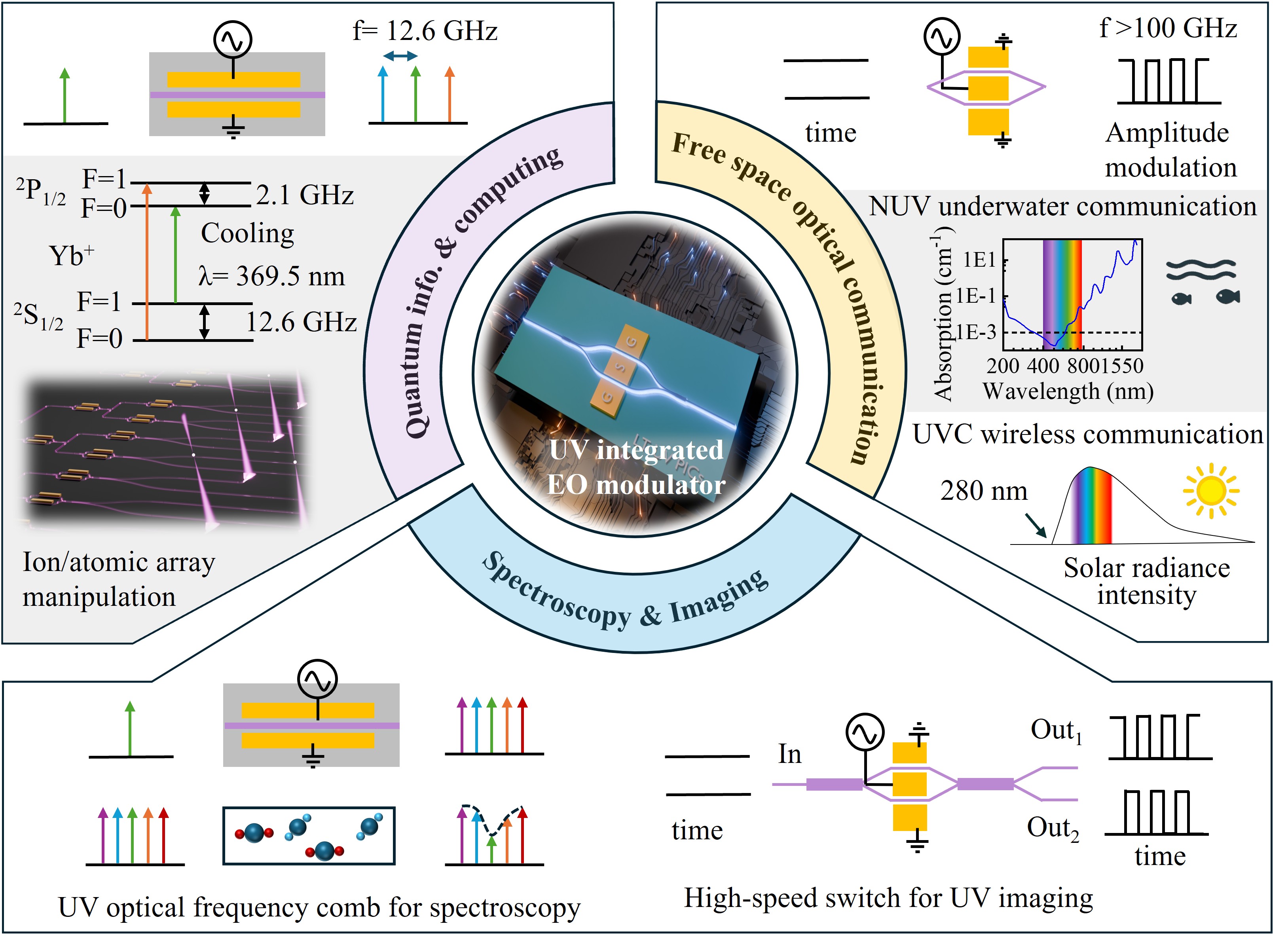}
\caption{Potential applications of high-speed UV integrated electro-optic modulator.}
\label{fig1}
\end{figure}

While DKDP crystals exhibit excellent UV transparency and a high electro-optic coefficient ($r_{63} \approx 25\text{ pm/V}$), their use is strictly limited to macroscopic bulk components. The inherent physical properties of this material, including its water solubility and mechanical fragility,  is fundamentally incompatible with standard thin-film deposition and nanofabrication processes, preventing its integration into scalable photonic circuits. Thin film lithium niobate (TFLN) has emerged as a workhorse for integrated EO modulator in the telecommunication bands due to its high Pockels coefficient of $r_{33} \approx 30\text{ pm/V}$ at 1550 nm. Recent advances have extended this performance into the visible spectrum. By optimizing traveling-wave electrode architectures, TFLN platforms have achieved a low half-wave voltage-length product ($V_\pi L$) of 0.17~V$\cdot$cm at 450~nm with 3-dB bandwidths exceeding 20 GHz \cite{Xue2023,Renaud2023}.  However, the fundamental bandgap of LiNbO\textsubscript{3} and the photorefractive effect, which causes optical damage and instability at high photon energies, severely hinder its application in the UV and deep-blue regimes. In contrast, thin-film lithium tantalate (TFLT) presents a compelling alternative for UV applications. While maintaining a high electro-optic coefficient comparable to TFLN, TFLT extends optical transparency down to 275~nm \cite{Juvalta2006,Wang2024}. Crucially, it exhibits a significantly suppressed photorefractive effect compared to its niobate counterpart, providing the requisite power stability and damage resistance for high-intensity UV photonic integration \cite{Kuznetsov2025}. While TFLT-based EO modulators have recently been demonstrated across the near-infrared and visible spectra \cite{Niels2026, Powell2025}, their extension into the UV remains an open challenge. This is primarily attributed to the dramatic increase in Rayleigh scattering, which scales as $1/\lambda^4$ and becomes the dominant loss mechanism at shorter wavelengths. Furthermore, while the sub-surface crystal defects induced by the hydrogen ion-slicing process are negligible in the visible/NIR regime, they likely act as absorption and scattering centers in the UV, where photon energies approach the electronic transitions of these defect states. Single mode LT waveguide shows propagation loss up to 4~dB/cm at a visible wavelength of 532~nm, which is much higher than that in NIR wavelength of 0.5~dB/cm waveguide \cite{Guo2026}. This rapid increase in attenuation trend suggests that achieving a low propagation loss of several~dB/cm remains challenging at UV wavelengths.

In this work, we demonstrate a high-performance integrated modulator based on a TFLT platform operating in the UV. We report the first realization of a single-mode LT waveguide at $\lambda$ = 375~nm with a propagation loss of 7.2~dB/cm. This modulator exhibits low insertion loss of 1.3~dB, ultra compact modulation length of 200~$\mu$m, low V\textsubscript{$\pi$}L of 85~mV$\cdot$cm and GHz modulation rate. To the best of our knowledge, this represents the first integrated UV electro-optic modulator to simultaneously achieve low insertion loss, compact device footprint, and gigahertz-scale modulation rates. This platform outperforms conventional bulk-crystal architectures, offering a two-order-of-magnitude reduction in $V_\pi L$ and a scalable path for complex UV photonic circuits. By providing a robust, high-speed solution for UV integrated photonics, this platform establishes a scalable architecture for the next generation of quantum information processors, ultra-precise atomic clocks, and high-bandwidth non-line-of-sight (NLOS) communication systems.

\section{Results}\label{sec2}
\subsection{Device design and implementation}\label{subsec2}

\begin{figure}[!t]
\centering
\includegraphics[width=1\linewidth]{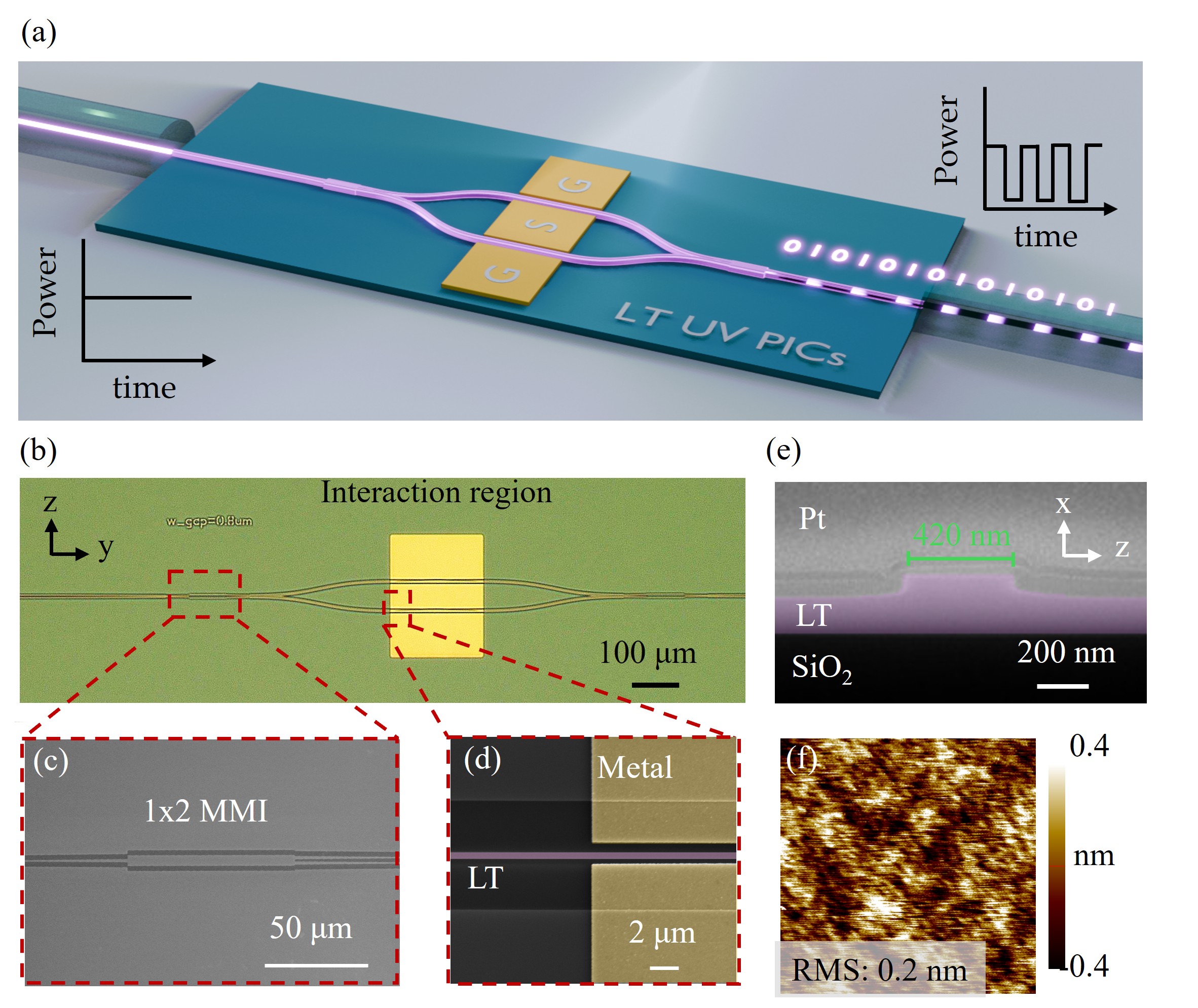}
\caption{Concept of LT UV modulator. (a) Schematic of integrated TFLT UV modulator. (b) Optical image of the fabricated device. (c) Scanning emission microscopy image of a 1x2 multi-mode interferometer. The length the of interaction region is 200~$\mu$m. (d) Scanning electron microscopy of image the electrode and LT waveguide. The gap between signal and ground electrodes is 1.5~$\mu$m. (e) Cross-section image of the LT waveguide with a tilt angle of 52 degrees. The arrows indicate the crystal axis of lithium tantalate. (f) Atomic force microscopy image of etched LT waveguide surface. The surface roughness RMS is measured to be 0.2~nm over the scanning area of 5x5~$\mu$m\textsuperscript{2}.}
\label{fig2}
\end{figure}

The strong electro-optic response of lithium tantalate enables efficient, high-speed phase modulation of the guided optical modes. While both microring resonators and Mach-Zehnder interferometers (MZIs) are widely used for integrated modulation, we employ an MZI architecture to ensure operational robustness. Unlike resonator-based modulators, which are highly sensitive to thermal fluctuations and fabrication errors, the MZI offers broadband operation and eliminates the need for complex, active resonance stabilization—a critical advantage for stable UV photonics. The architecture of our integrated TFLT UV modulator is schematically illustrated in Fig. 2(a). The MZI features a dual-arm phase modulator in a push-pull configuration driven by a ground-signal-ground (GSG) lumped electrode, where the applied electric field induces refractive index changes of opposite signs in each arm. This differential phase shift effectively doubles the total phase accumulation for a given bias, thereby halving the half-wave voltage ($V_\pi$) and enhancing the modulation efficiency. In this work, a 375~nm UV source (Bios375-20FC, Elite optoelectronics co. ltd) is coupled via a cleaved single-mode fiber (SM300, Thorlabs) into the polished LT chip. The LT waveguide facets are tapered to 2.7~$\mu$m to minimize modal mismatch and enhance coupling efficiency to a simulated fiber-to-waveguide loss of 7.5~dB per facet. A $1 \times 2$ multi-mode interferometer (MMI) splits the light equally into two arms. Both beams are then modulated by the phase shifter before interfering at a second, symmetric $1 \times 2$ MMI. The resulting signal is coupled back into a single-mode fiber. The combination of a high EO coefficient and a short working wavelength enables an exceptionally compact device footprint. A modulation length of only 200~$\mu$m is sufficient to reach a $2\pi$ phase shift at a CMOS-compatible driving voltage of 4.2V. This ultra-compact interaction length is two orders of magnitude smaller than those typically required at telecommunication wavelengths. This disparity highlights the inherent scaling advantage of the UV regime: because the induced phase shift ($\Delta\phi = 2\pi L \Delta n / \lambda$) is inversely proportional to wavelength, the modulation effect is naturally magnified as $\lambda$ decreases. The length of the modulator is 1.16 mm (Fig. 2(b)), representing a highly footprint-efficient solution for UV integrated photonics compared to traditional bulk components. Scanning electron microscopy (SEM) images in Fig. 2(c) and 1(d) provide magnified views of the MMI and the phase modulator input, respectively. A narrow electrode gap of 1.5~$\mu$m was used to maximize the electric field strength and modulation efficiency, while minimizing metal-induced optical losses. The PICs were patterned on a commercial 300~nm-thick x-cut LT-on-insulator platform with a 100~nm etch depth. The TFLT platform provides high index contrast, which is essential for tight mode confinement and the realization of small-footprint UV circuits. To ensure single-mode operation, critical for maintaining high extinction ratios (ER) in the MZI, the waveguide width is tapered from 2.70~$\mu$m to 0.42~$\mu$m. Figure 2(e) shows a SEM cross-section of the waveguide. A platinum layer was deposited prior to imaging to mitigate charging effects caused by the insulating substrate. Vertical sidewalls were achieved via wet chemical etching using a mixture of $\text{H}_2\text{O}_2 / \text{KOH} / \text{C}_6\text{H}_8\text{O}_7$\cite{Nenezic2025}. Minimizing surface roughness is critical to reducing scattering and propagation losses, which scale as $1/\lambda^4$ and are therefore significantly more detrimental at 375~nm than at infrared wavelengths. Atomic force microscopy (AFM), reported in Fig. 2(f), measurements confirmed a smooth etched surface with a root-mean-square (RMS) roughness of 0.2~nm over a 5$ \times$5~$\mu$m\textsuperscript{2} area, demonstrating the high precision of our fabrication process in achieving sub-nanometer surface quality.

\subsection{Passive device characterization}\label{sec3}

\begin{figure}[!t]
\centering
\includegraphics[width=1\linewidth]{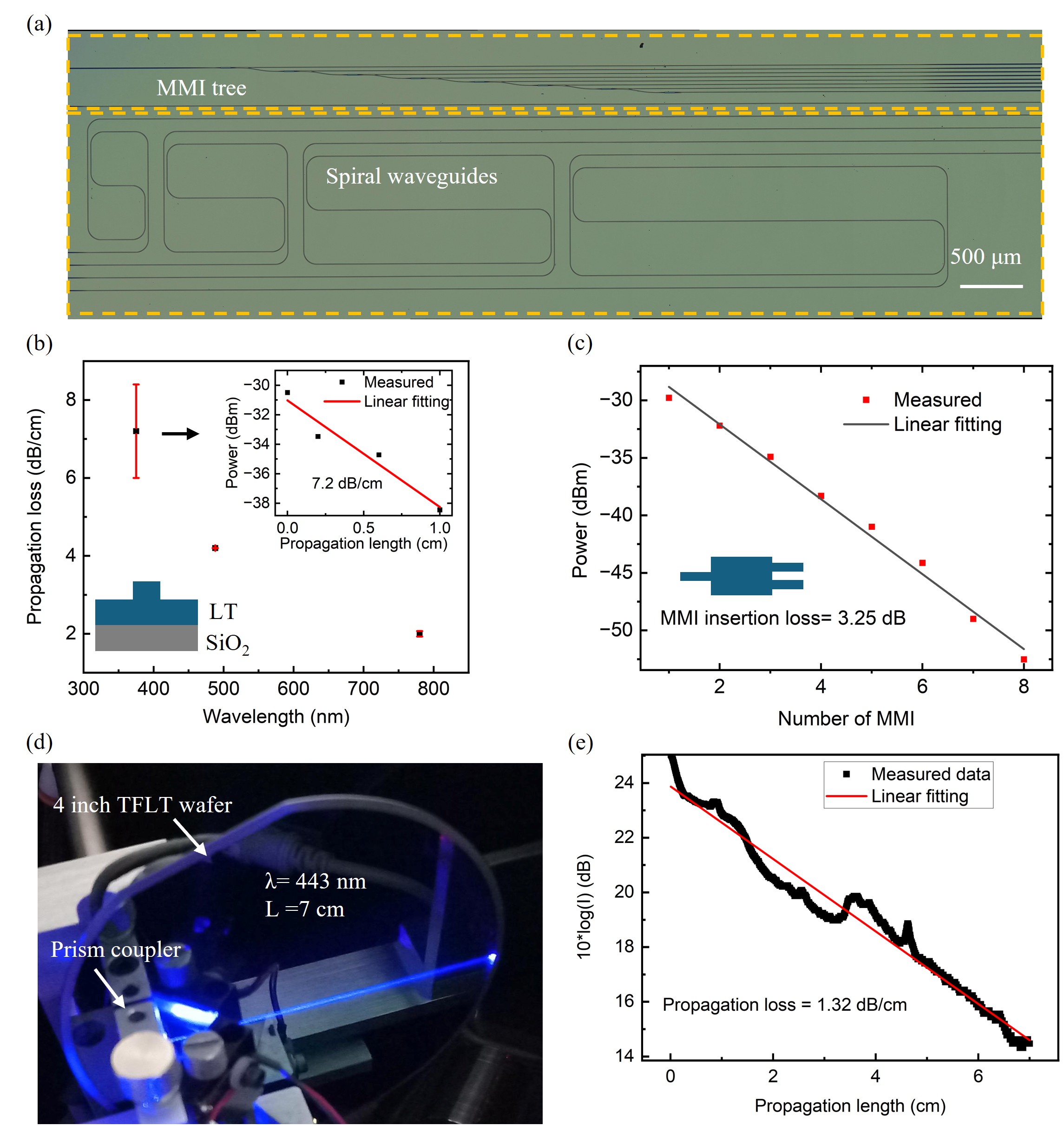}
\caption{TFLT components characterization. (a) Optical image of the fabricated chip with MMI trees and spiral waveguides for loss measurement. (b) Relationship between the propagation loss of the LT waveguides and operating wavelengths. The inset shows the linear fitting of measured intensity over propagation length at a wavelength of 375~nm. (c) The relationship between the intensity and the number of MMIs. (d) Photo of light path in a 600~nm thick TFLT wafer at a wavelength of 443~nm. (e) Relationship between the scattered intensity (I) along the propagation length, obtained from (d). The loss of the thin film is estimated to be 1.32~dB/cm at 443~nm.
}
\label{fig3}
\end{figure}

The passive optical performance of the device was evaluated through a comprehensive loss characterization. The propagation loss was characterized using a set of spirals of different lengths. To mitigate the effects of anisotropic wet etching, where lateral etch rates depend on the crystal plane, pre-compensation is applied to the width of spiral characterization waveguides to reduce the bend loss due to the width variation. The device architecture is optimized for the TE mode, with propagation along the y-axis to leverage the crystal’s $r_{33}$ coefficient (aligned with the z-axis, see Fig. 3(a)). This orientation maximizes the electro-optic overlap, thus enhancing the modulation efficiency. To eliminate systematic errors associated with mode mismatch loss in the curved sections, the geometry of each spiral was constrained to include the same number of bends. This design ensures that the power variation between different spirals is solely a function of the propagation distance along the y-axis, providing a reliable loss measurement. Figure 3(b) illustrates the propagation loss as a function of wavelength, spanning from the UV to the visible spectrum. The inset displays the linear fitting used to extract the loss at 375~nm, confirming a loss of 7.2~dB/cm for a single-mode LT waveguide in this regime. While the propagation loss increases rapidly from a near-infrared wavelength of 780~nm (2~dB/cm) to the UV due to enhanced scattering, the associated device footprint scales down accordingly. The 200-$\mu$m-long phase modulator leads to a total insertion loss of only 0.14~dB in the UV. This value is comparable to the losses observed at visible or NIR wavelengths, where interaction lengths on the order of millimeters to centimeters are typically required to achieve low-voltage operation. Consequently, the TFLT platform maintains high optical efficiency even at the high-scattering UV regime. To characterize the performance of the 1$\times $2 MMI splitters, an MMI tree structure is utilized to extract the insertion loss through a linear fit of the output power versus the number of cascaded stages (Fig. 3(c)). The measured insertion loss of 3.25~dB per MMI, which includes the intrinsic 3~dB splitting ratio, aligns well with the simulated value of 3.11~dB. Taking into account the insertion loss of MMI and propagation loss of the waveguide, the modulator with a device length of 1.16 mm shows an insertion loss of 1.3 dB. The intrinsic optical quality of the TFLT layer was investigated via prism-coupling on a 600-nm thick TFLT wafer. By exciting the fundamental TE slab mode, a visible propagation path exceeding 7~cm can be observed in Fig. 3(d). Figure 3(e) displays the log-scale intensity of the scattered light as a function of the propagation length. The linear fit of this experimental data reveals an attenuation of 1.32~dB/cm at 443~nm. With a measured surface roughness below 0.5~nm, the majority of the observed optical attenuation is attributed to the fundamental properties of the thin-film lithium tantalate. The propagation loss of 1.32~dB/cm at 443~nm thus represents the intrinsic absorption floor of the current TFLT wafer, providing a critical benchmark for the material performance limits as we transition into the UV. TFLT waveguide can be further improved by annealing in an oxygen atmosphere at 520~$^oC$ to repair crystal defects and reduce material absorption after processing, which is not performed in this work but will be investigated in future work \cite{ Shams-Ansari2022}.

\subsection{Electro-optical performance of UV LT modulator}\label{sec3}

\begin{figure}[!t]
\centering
\includegraphics[width=1\linewidth]{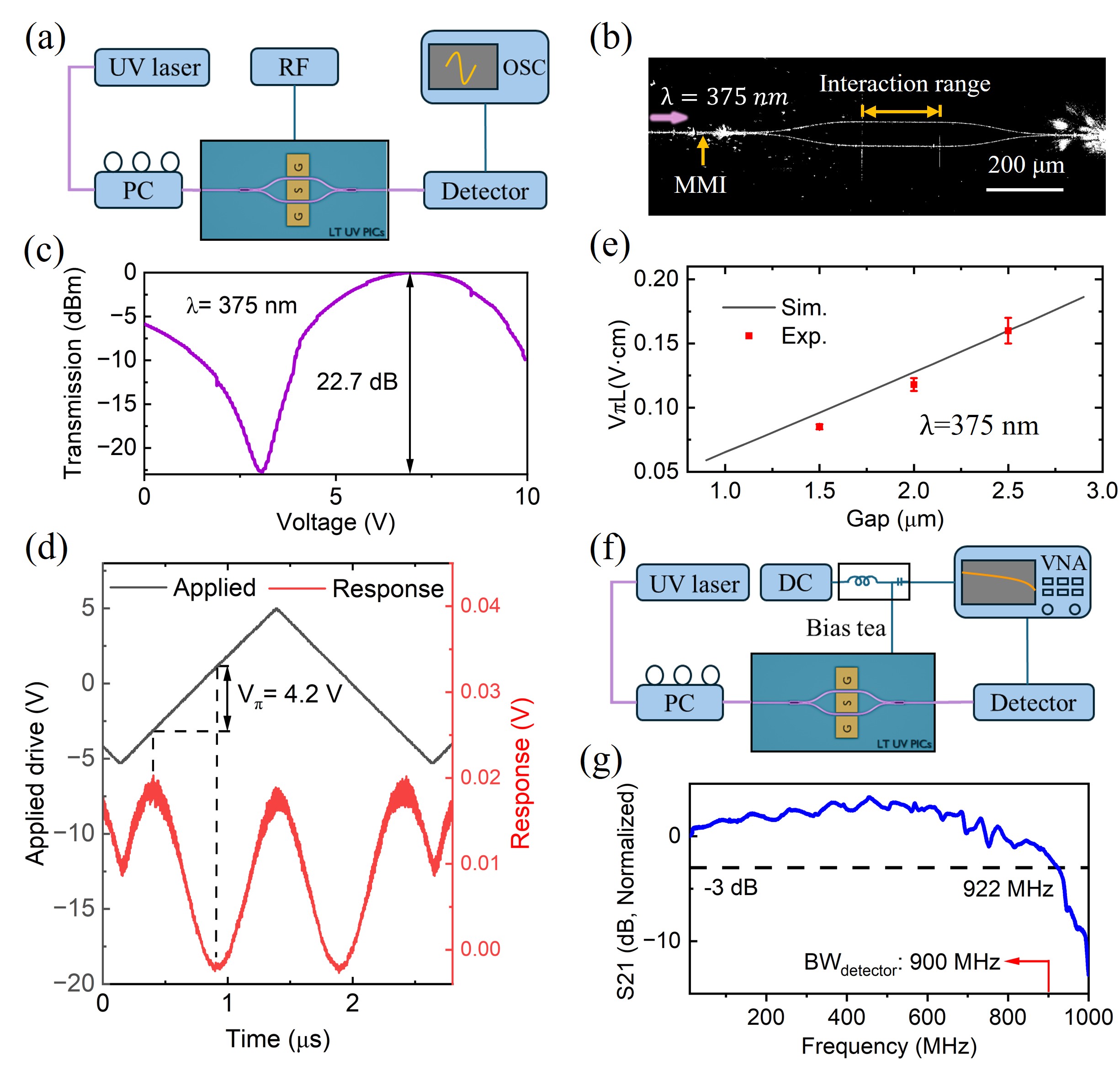}
\caption{
EOE response of the TFLT UV modulator. (a) Measurement setup used for V$_\pi$ measurement. (b) Top microscope image of light at 375~nm propagating in the UV modulator. (c) Normalized transmitted power at the output port of the modulator as a function of the applied voltage. An extinction ratio of 22.7~dB is obtained. (d) Relationship between the V$_\pi$L and electrode gap in simulation and experiments at a wavelength of 375~nm. (e) The relationship between the applied triangular signal and the modulator response. The triangular wave has a frequency of 400 kHz and an amplitude of 10V. The V$_\pi$ of 4.2 V is measured in push-and-pull configuration with a ground-signal-ground electrode. (f) Measurement setup to obtain the EOE frequency response measurement of the modulator. (g) Frequency response of EOE link consisting of the UV TFLT modulator and an APD photodiode with a 3-dB bandwidth of 900 MHz\cite{APD410}.}
\label{fig4}
\end{figure}

The electro‑optic performance of the modulator was evaluated by measuring the half‑wave voltage $V_{\pi}$ and bandwidth. Figure 4(a) shows the measurement setup diagram used for characterizing the half-wave voltage of the modulator. A UV diode laser of 375~nm is coupled to a single mode fiber with an output power of 7 mW. A fiber polarization controller (FPC) tunes the polarization of light in the fiber to enable transverse electric (TE) mode excitation in the LT waveguide. As the phase modulator is configured to modulate TE mode only, any residual light of transverse magnetic (TM) polarization lowers the extinction ratio of the modulator. The output is collected by a single mode fiber and coupled to an avalanche photodiode (ADP, APD410, Menlo Systems) for the response measurement of the modulator. Figure 4(b) shows the scattered light from LT photonic integrated circuits at an operating wavelength of 375~nm. The absence of visible intensity decay along the propagation indicates the low insertion loss of the modulator. When conducting the measurement of the extinction ratio of modulators, a powermeter (S120VC, Thorlabs) is used to measure the optical power when the voltage linearly increases from 0 to 10 V. The measured extinction ratio is as high as 22.7~dB as shown in Fig. 4(c), suggesting a highly robust fabrication process. 

To investigate V$_\pi$ of each modulator, a triangular signal of 400 kHz is applied on the electrodes to eliminate the impact of thermal and direct current (DC) drift. Figure 4(d) shows the relation among the applied signal, modulator response and time. The V$_\pi$ of the modulator was determined to be 4.2 V, calculated by measuring the voltage swing required to transition the optical response between its minimum and maximum transmission points. Given an electrode length of 200 um, this corresponds to a low V$_\pi$L of 85 $\pm$ 2~mV$\cdot$cm. Measurements of $V_{\pi}L$ across six modulators yielded a standard deviation of 2~mV$\cdot$cm, reflecting excellent device uniformity and fabrication process reproducibility. These experimental results matches well with the simulation, with V\textsubscript{$_\pi$}L directly proportional to the gap width between metal electrodes, as shown in Fig. 4(e). Such a low V\textsubscript{$_\pi$}L of 85~$\pm$ 2~mV$\cdot$cm is one to two orders of magnitude lower than that in telecom wavelengths. This improvement is attributed to the shorter operating wavelength, higher material EO coefficient and stronger mode confinement in the UV. This experimental value is slightly lower than the simulated value of 96~$\pm$2~mV$\cdot$cm. This discrepancy may result from deviations between the simulated and fabricated structures, including a larger effective Pockels coefficient than the value assumed in the simulations, a reduced electrode gap in the fabricated device, or a combination of these factors. A Pockels coefficient $r33$ of 35 pm/V is used in the simulation at the operating wavelength of 375~nm \cite{Juvalta2006}.

The dramatic reduction in V$_\pi$L enables operation with an interaction length as short as 200 $\mu m$ for a 4.25 V drive voltage. This short interaction length, in turn, reduces on‑chip insertion loss while significantly simplifying the design of high-speed electrodes because the velocity matching requirement is drastically reduced. At these ultra-short scales, the velocity mismatch between the optical and the microwave mode becomes negligible, enabling efficient, high-speed modulation without the dispersion penalties typically encountered in centimeter-scale NIR devices. Due to the limited responsivity of UV-sensitive photodiodes and the moderate power levels of available laser sources, an avalanche photodiode with an integrated transimpedance amplifier (TIA) was used to ensure a robust signal for the $S_{21}$ electrical-to-optical-to-electrical (EOE) response characterization. The experimental setup for the high-speed characterization is detailed in Fig. 4(f). A Vector Network Analyzer (VNA) is used to measure the electro-optic frequency response of the TFLT modulator at a wavelength of 375~nm. However, as the 3-dB bandwidth of the avalanche photodiode is manufacturer-rated at 900 MHz\cite{APD410}, the measured system response is currently detector-limited. Figure 4(g) shows the EOE response of the complete optical link (including both the modulator and the detector). A measured 3-dB bandwidth of 922 MHz aligns well with the performance of the photodetector. Due to the lack of commercially available high‑speed UV modulators for photodetector calibration and the absence of relevant manufacturer specifications, subtraction of the detector response could not be performed.

\begin{figure}[!t]
\centering
\includegraphics[width=1\linewidth]{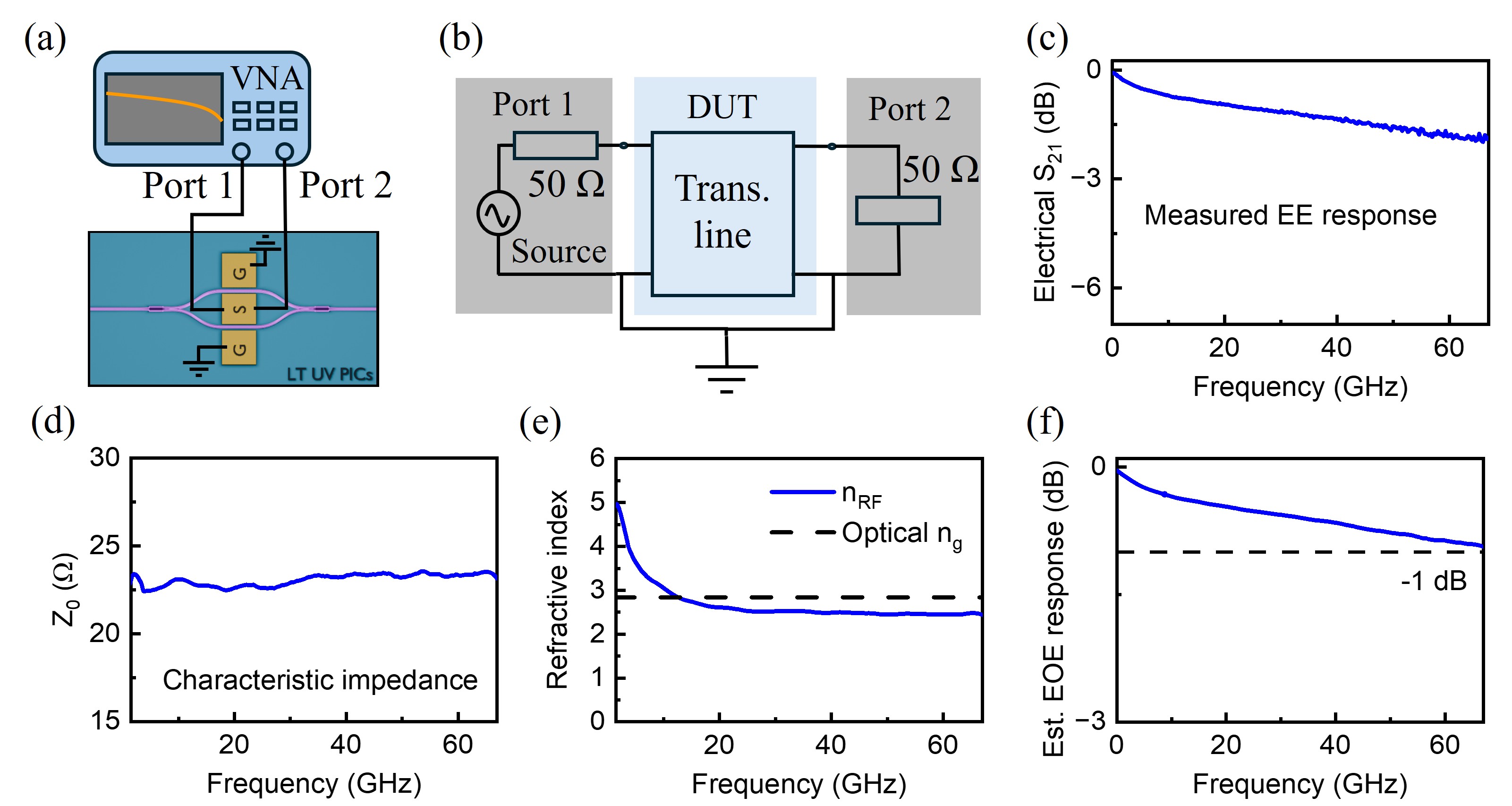}
\caption{EE response of TFLT UV modulator. (a) Measurement setup to obtain the EE response. (b) Schematic model of the measurement setup in (a). (c) Measured $S_{21}$ transmission response up to 67 GHz. (d)Extracted characteristic impedance ($[Z_0]$) as a function of frequency.(e) Effective microwave refractive index $n_{RF}$ versus frequency; the dashed line indicates the calculated optical group index $n_g$ of the UV guided mode,  (f) Estimated intrinsic EOE response. The blue and black solid curves isolate the contributions from RF loss and velocity mismatch, respectively. }
\label{fig5}
\end{figure}

Although the EOE characterization is currently constrained by the available equipments, the electrical-to-electrical (EE) response of the electrodes were characterized using a VNA up to 67 GHz (Keysight N5247B) to evaluate the intrinsic microwave properties of the TFLT UV modulator. The measurement configuration and the corresponding simplified equivalent circuit are illustrated in Fig. 5(a) and (b), respectively. As shown in Fig. 5(c), the measured $S_{21}$ transmission response exhibits a low microwave attenuation of only 2~dB at 67~GHz. To further explore the electrode properties, the characteristic impedance, microwave refractive index $n_{RF}$ were extracted from the measured S-parameters to estimate EOE response(Fig. 5(d–f)). Because of the quasi-TEM nature of coplanar waveguide electrodes, the characteristic impedance is a stable 23~$\Omega$ across the frequency range (Fig. 5(d)). This indicates that the modulation bandwidth could be further optimized by terminating the electrodes with a 23~$\Omega$ load to exclude any reflections at the open end of the electrodes. While high-speed modulation in traveling-wave architectures typically requires rigorous phase matching between microwave and optical fields, this device operates in a regime where the effective microwave index of 2.3 and optical group index of 2.8 are close to each other (Fig. 5(e)). This index control remains a critical asset for scaling to longer traveling-wave phase modulators, which are essential for applications such as UV frequency comb generation. Given the ultra-compact 200~$\mu$m interaction length of this modulator, the velocity mismatch is effectively negligible, as corroborated by the estimated EOE response in Fig. 5(f). This EOE response is estimated by taking into account the RF loss, the velocity mismatch between the optical mode and microwave mode, and the impedance matching by terminating with 23 Ohm to minimize reflections, based on the measured EE response \cite{Witzens2018}. The estimated EOE roll-off is 1~dB at a microwave frequency of 67 GHz, far below the 3~dB threshold. These results confirm that while the experimental EOE bandwidth is currently capped at 922~MHz due to detector limitations, the underlying TFLT architecture is intrinsically capable of supporting modulation bandwidths into the multi-GHz regime.

\subsection{Discussion}

\begin{figure}[!t]
\centering
\includegraphics[width=1\linewidth]{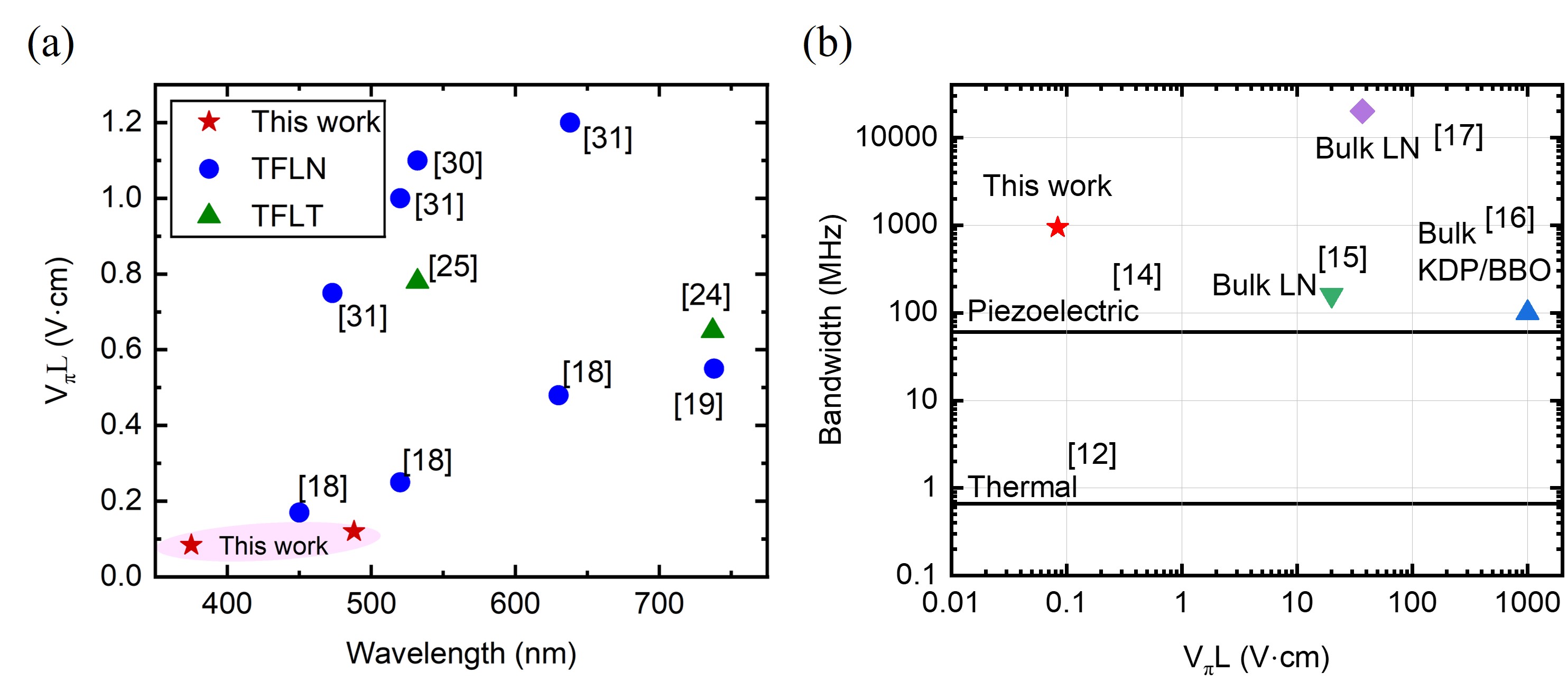}
\caption{Performance benchmarks for state-of-the-art integrated electro-optic modulators in UV and visible wavelengths. (a) $V_{\pi}L$ as a function of operating wavelength across the UV and visible spectra, reported in this work and state-of-the-art integrated modulators\cite{Xue2023,Renaud2023,Powell2025,Guo2026,Li2022,Hara2024}. (b) Comparison among this work, reported on-chip devices\cite{Wang2026,Castillo2026} and commercial bulk devices\cite{qubig_pm11_uv,qubig_am8_uv,qubig_pc_uv} operating at UV wavelengths, in terms of modulation bandwidth and $V_{\pi}L$.}
\label{fig6}
\end{figure}

Our integrated UV EO modulator represents a significant leap over state-of-the-art platforms in terms of footprint, modulation efficiency, and high-speed potential. As illustrated in Fig. 6, we compare the V$_\pi$L of our TFLT modulator against existing integrated modulators spanning the visible and UV regimes \cite{Li2022,Hara2024}. TFLN has emerged as a frontrunner for integrated EO modulators in the NIR and visible regimes, achieving a $V_\pi L$ of $0.17\text{ V}\cdot\text{cm}$ at $450\text{~nm}$ \cite{Xue2023}. However, as the operational wavelength shifts toward the UV, TFLN faces severe limitations due to the photorefractive effect \cite{Xu2021,Guo2026}. This effect scales aggressively at shorter wavelengths and higher optical powers, fundamentally hindering the reliability of TFLN for UV applications. TFLT has recently been shown to support watt-level high-power handling at 775~nm \cite{Kuznetsov2025}. While TFLT electro-optic modulators have previously demonstrated high-speed performance at telecom and green ($532\text{~nm}$) wavelengths, our work realizes the first integrated EO modulator operating in the UV at 375~nm (Fig. 6(a)). We report a record-low $V_\pi L$ of 0.085~V$\cdot$cm at 375~nm and 0.12~V$\cdot$cm at 448~nm. To the best of our knowledge, these values represent the highest modulation efficiencies achieved in the UV-visible spectrum to date. A comparative analysis of the integrated UV modulation landscape is presented in Fig. 6(b). While thermo-optic (TO) modulators offer high integration density and broad spectral coverage due to their simplistic architecture, their operational speeds are fundamentally capped at the tens of kHz regime by thermal diffusion times \cite{Wang2026}. More recently, piezoelectric-based intensity modulators on AlN-on-AlO\textsubscript{x} platforms have demonstrated a significant leap in performance, pushing bandwidths into the tens of MHz range. While electro-optic modulation at tens of GHz is a cornerstone of modern NIR telecommunications, extending this performance to the UV has remained challenging. Traditional UV-transparent crystals such as BBO and KDP are currently restricted to bulky geometries that are incompatible with wafer-scale fabrication. Their weak light-matter interaction results in prohibitive V$_\pi$L values in the kV$\cdot$cm range and bandwidths capped at 100 MHz. In contrast, bulk lithium niobate modulators leverage a superior EO coefficient to achieve V$_\pi$L values in the tens of~V$\cdot$cm. However, their intensity modulation bandwidth is frequently thermal-limited to $\sim$160 MHz due to the high RF drive powers required for bulk electrodes. Furthermore, the photorefractive effect in LN constrains the input power density to below $0.5\text{ W/mm}^2$ (AM8-UV, QuBIG Inc.), severely limiting its utility in high-brightness UV applications.

In this work, the integrated LT modulator demonstrated a low V$_\pi$L, a high bandwidth of sub-GHz, and a low insertion loss of 1.3~dB. Crucially, the platform exhibits optical stability. No photodarkening was observed even at an on-chip optical intensity of 5 kW/mm\textsuperscript{2}, calculated from a 700 $\mu$W guide power confined within a sub-wavelength mode volume of 0.14~$\mu$m\textsuperscript{2}. To contextualize the performance of this device, we compare our results with other previously demonstrated on-chip and commercial bulk UV modulators in terms of bandwidth and V$_\pi$L (banwidth/V$_\pi$L ratio) in Fig. 6(b). By this metric, our TFLT modulator demonstrates a two-order-of-magnitude improvement over the bulk lithium niobate modulator and a four-order-of-magnitude improvement over the traditional bulk modulator made of BBO/KDP crystals. 

Furthermore, this work realizes the first UV photonic integrated circuit based on the thin-film lithium tantalate platform. While purely passive platforms such as AlO\textsubscript{x} currently offer lower attenuation ($<$dB/cm) in the UV \cite{Lin2022,Neutens2025}, they lack the intrinsic EO properties required for ultra-fast signal processing. By contrast, the high EO coefficient of TFLT enables a unique synergy of high-speed modulation and potential frequency conversion that is unattainable in passive thin-film oxides. Heterogeneous integration of TFLT on the passive platform via micro-transfer printing is a promising approach to combine the advances of TFLT and AlO\textsubscript{x} platforms. This method has been demonstrated to enrich the functionality of the passive SiN platform at telecom wavelengths\cite{Niels2026}. However, it still remains challenging to integrate LT on AlO\textsubscript{x} because the large index contrast between TFLT and AlO\textsubscript{x} requires a much smaller taper to ensure a low-loss transition between AlO\textsubscript{x} and LT waveguides. In this work, by establishing a new paradigm for the integration of active UV optoelectronic devices using the TFLT platform, the long-standing challenges of device miniaturization and power-handling stability in the UV spectrum have been addressed. The reported V$_\pi$L of 85~mV$\cdot$cm and the sub-GHz bandwidth capability provide a scalable architecture for wafer-scale UV integrated photonics. These results enable the transition from lab-scale bulk optical setups to robust, chip-scale systems, including on-chip trapped-ion quantum computing, portable atomic clocks, UV spectroscopy, and high-bandwidth solar-blind UV communications.

\section{Methods}
\subsection{Fabrication}
The UV photonic integrated circuits are patterned on top of a commercial x-cut 300~nm thick thin film lithium tantalate ordered from NanoLN Inc. A 200~nm thick SiO\textsubscript{2} layer is deposited on top of the TFLT as a hard mask using inductively coupled plasma-enhanced chemical vapor deposition (ICP-PECVD). The photonic circuits are firstly patterned on the resist (ARP6200.13) via electron beam lithography (EBL) and then transferred to the SiO\textsubscript{2} layer by reactive ion etching (RIE). The etching of lithium tantalate is realized by wet chemical etching in a mixture solution of citric acid/KOH/H\textsubscript{2}O\textsubscript{2} at a temperature of 85~$^oC$. Finally, the SiO\textsubscript{2} hard mask is removed in buffered Hydrogen fluoride (HF) solution to complete the fabrication of LT waveguides. 

The electrodes are produced by a lift-off process. The 800~nm thick resist (ARP6200.18) is patterned by EBL. After depositing a 220~nm-thick Ti/Au metal stack via electron beam physical vapor deposition (EB-PVD), the sample is immersed in ARP600-71 solution to implement the lift-off process. The sample is then diced, and the side facets are polished to maximize the coupling efficiency.

\subsection{Measurement}

The optical quality of TFLT is estimated by measuring the absorption loss of 600~nm thick four-inch TFLT wafer via the Metricon 2010/M prism coupling setup. The light is coupled to the fundamental TE mode of the slab mode in the thin film. The intensity of the scattered light is measured via a detector that scans along the propagation path and records the intensity of scattered light. As the top surface of the thin film has a negligible surface roughness of 0.2~nm, the measured loss is mainly attributed to the absorption of the layer. The loss is obtained by implementing linear fitting of the log scale intensity over the propagation length, as shown in Fig. 3(e).

\section{Acknowledgments}
This work is supported by  Horizon Europe Project Qu-PIC (101135845).

\section{Authors contributions}
C.L. conceived, performed, and analyzed the experiments. C.L. designed and fabricated the device. C.L. and T. V. simulate the V$_\pi$ of the device. C.L., P.N., M.N, V.B.O and M.B. developed the waveguide process flow, K. A. measured the TFLT absorption loss, C.L., S.A., H.L. and A.M. measured the high-speed response of the device. The project was supervised by B.K.. All authors reviewed and approved the manuscript.

\section{Competing interests}
The authors declare no competing interests.

\section{Data availability}
The datasets generated and analyzed during the current study are available
from the corresponding authors upon reasonable request.

\bibliography{TFLT}

\end{document}